\documentclass[aps,prl,twocolumn,nofootinbib,preprintnumbers,10pt]{revtex4-1}

\pdfoutput=1
\usepackage{amssymb,amsmath,amsthm,graphicx,ulem}
\usepackage{hyperref}
\usepackage{graphicx,subfigure}
\usepackage{ulem}
\usepackage{epsfig}
\usepackage{amsmath}
\usepackage{amsfonts}
\usepackage{amssymb}
\usepackage[usenames]{color}
\usepackage[letterpaper,left=2.cm,right=2.cm,top=2.5cm,bottom=2.5cm]{geometry}
\usepackage[T1]{fontenc}
\usepackage{rotating}

\newcommand{\beq}{\begin{equation}}
\newcommand{\eeq}{\end{equation}}
\newcommand{\be}{\begin{equation}}
\newcommand{\ee}{\end{equation}}
\newcommand{\beqa}{\begin{eqnarray}}
\newcommand{\eeqa}{\end{eqnarray}}
\newcommand{\beqar}{\begin{eqnarray*}}
\newcommand{\eeqar}{\end{eqnarray*}}
\newcommand{\bea}{\begin{eqnarray}}
\newcommand{\eea}{\end{eqnarray}}
\newcommand{\f}[2]{\frac{#1}{#2}}

\newcommand{\p}{\partial}

\def\cO{{\cal O}}
\def\cS{{\cal S}}

\newcommand{\nn}\nonumber

\begin{document}

\title{Evolution of nonlocal observables in an expanding boost-invariant plasma}

\author{Juan F. Pedraza}
\affiliation{\vskip .1cm Theory Group, Department of Physics and Texas Cosmology Center, The University of Texas at Austin, Austin, TX, 78712, USA.}

\preprint{UTTG-14-14, TCC-014-14}

\begin{abstract}
Using the AdS/CFT correspondence, we compute analytically the late-time behavior of two-point functions, Wilson loops and entanglement entropy in a strongly coupled $\mathcal{N}=4$ super-Yang-Mills plasma undergoing a boost-invariant expansion. We take into account the effects of first order dissipative hydrodynamics and investigate the effects of the (time-dependent) shear viscosity on the various observables. The two-point functions decay exponentially at late times and are unaffected by the viscosity if the points are separated along the transverse directions. For longitudinal separation we find a much richer structure. In this case the exponential is modulated by a nonmonotonic function of the rapidities and a dimensionless combination of the shear viscosity and proper time. We show that this peculiar behavior constrains the regime of validity of the hydrodynamical expansion. In addition, similar results are found for certain Wilson loops and entanglement entropies.
\end{abstract}

\maketitle

\noindent \textbf{1. Introduction.} Studying real-time phenomena in quantum chromodynamics (QCD)
is a difficult task and a major focus of current research. Besides the pure theoretical motivation,
understanding QCD in out-of-equilibrium scenarios is
relevant for experimental programs of heavy-ion collisions at RHIC and LHC. According to the current paradigm, the colliding matter creates a soup of deconfined quarks and gluons that thermalizes fast, expands and finally hadronizes. At the relevant energies achieved in these experiments, QCD is strongly coupled and standard perturbative techniques are inadequate, creating
a demand for new theoretical tools. In recent years, the discovery of the AdS/CFT correspondence \cite{AdSCFT}
has granted us access to the study of a large class of strongly coupled non-Abelian
gauge theories in a complete nonperturbative way.

In the last stage of the evolution, various observables of the quark-gluon plasma are well described in terms of hydrodynamics \cite{Hydro} and, to a good extent, it seems that it behaves approximately as a perfect fluid \cite{Bjorken}.
In the context of AdS/CFT, the first steps towards understanding the late-time expansion of the plasma were given in \cite{BIPlasma}, assuming boost invariance along the collision axis (which is believed to hold in the mid-rapidity region \cite{Bjorken}) and homogeneity/isotropy in the transverse plane. Of course, these simplifying assumptions were vital for having an analytical handle on the problem but one ultimately wants to relax these conditions in order to model a more realistic plasma, e.g. including anisotropic effects generated by off-center collisions \cite{Anisotropic} and radial flow due to finite size nuclei \cite{Radial}. Going beyond the hydrodynamical description requires one to overcome several challenges, as it requires a full numerical solution to the initial value problem in asymptotically AdS spaces \cite{Numerical}.

One way to characterize the evolution of the plasma in a time-dependent setup is by studying the behavior of nonlocal observables and analyzing the way in which they reach equilibrium. Indeed, this approach has been used with great success in the program of holographic thermalization \cite{thermalization}, where the system is excited by the injection of a spatially uniform density of energy and eventually equilibrates. On the gravity side, such quenches are described in terms of the gravitational collapse of a shell of matter that leads to the formation of a black hole. These are toy models that describe the early-time evolution of the plasma, before entering the regime of expansion and cooling. The idea here is to extend these results to another regime of interest that is analytically tractable, namely the stage in which the plasma is described by hydrodynamics.

In the framework of nonlinear dissipative hydrodynamics, the fluid/gravity duality \cite{Fluid} provides us with full control of the bulk geometry in a wide variety of scenarios. For concreteness, we will assume boost invariance along the axis of expansion and translational/rotational
symmetry in the transverse plane as in \cite{BIPlasma}. We will phrase our discussion in terms of the simplest theory with a known gravity dual, i.e. $\mathcal{N}=4$ supersymmetric Yang-Mills at large $N_c$ and 't Hooft coupling $\lambda$ but, appealing to universality, we expect our results to hold under more general circumstances.

\noindent \textbf{2. Holographic model.}
When assuming boost invariance, it is natural to work in proper time and spacetime rapidity coordinates $(\tau,y,x_1,x_2)$, where
\be
t=\tau \cosh y\,,\qquad x_3=\tau \sinh y\,.
\ee
In Fefferman-Graham coordinates, the most general bulk
metric (for the given symmetries) takes the form
\be\label{metric}
ds^2=\f{1}{z^2} \left( -e^{a} d\tau^2+e^{b} \tau^2
dy^2 +e^{c} d\vec{x}_\perp^2 +dz^2\right)\,,
\ee
where the coefficients $a$, $b$ and $c$ depend on $z$ and $\tau$. In order to study the large proper time limit of the metric we define the scaling variable $v\equiv z\tau^{-\f{1}{3}}\epsilon^{\f{1}{4}}$ (where $\epsilon$ is a dimensionful constant), and take the limit $\tau \to \infty$ with $v$ fixed. Then, we expand the coefficients $a$,
$b$ and $c$ as series of the form
\be
\label{e.aexp}
a(z,\tau)=a_0(v)+a_1(v) \tau^{-\f{2}{3}} +a_2(v)
\tau^{-\f{4}{3}} +\ldots\,,
\ee
with similar expressions for $b(z,\tau)$ and $c(z,\tau)$. The zeroth and first order coefficients were computed in \cite{BIPlasma} by solving the vacuum Einstein equations perturbatively. They are given by
\bea
&&a_0 = \log\f{(1-\f{v^4}{3})^2}{1+\f{v^4}{3}}\,,\quad \tilde{b}_0=3c_0 = \log(1+\tfrac{v^4}{3})^3\,,\nonumber\\
&&a_1 =  \f{2\eta_0(9+v^4)v^4}{9-v^8}\,,\quad \tilde{b}_1 = - \f{6 \eta_0 v^4}{3+v^4}\,,\\
&&c_1 = - \f{2\eta_0v^4}{3+v^4} -\eta_0 \log \f{3-v^4}{3+v^4}\,,\nonumber
\eea
where $\tilde{b}_i(v) \equiv b_i(v)+2c_i(v)$ and $\eta_0=2^{-\f{1}{2}}3^{-\f{3}{4}}\epsilon^{-\f{1}{4}}$. The metric (\ref{metric}) is dual to a plasma that is expanding along the $x_3$ direction, with the transverse plane spanned by $\vec{x}_\perp=\{x_1,x_2\}$.
Given this geometry, we would like to study the evolution of nonlocal observables such as two-point functions, Wilson loops and entanglement entropy. From the bulk point of view, this problem amounts to the computation of certain extremal surfaces with fixed boundary conditions \cite{thermalization}.

To analyze the physical content of the metric (\ref{metric}) in terms of the boundary theory, we can compute the expectation value of the energy-momentum tensor using the standard techniques of holographic renormalization \cite{renormalization}. The only nonvanishing components are $T_{\tau\tau}$, $T_{yy}$ and $T_{xx}\equiv T_{x_1x_1}=T_{x_2x_2}$. More specifically, at zeroth order one finds $T^{\mu\nu}=(\varepsilon+p)u^\mu u^\nu-p g^{\mu\nu}$, with $\varepsilon=3p=\tau^{-\frac{4}{3}}\epsilon$, i.e. a perfect fluid energy-momentum tensor, with conformal equation of state, satisfying the so-called Bjorken hydrodynamics \cite{Bjorken}. 
At next order, there is an additional dissipative term in the energy-momentum tensor, which is of first order in gradients: $\tau^{\mu\nu}=-\eta(\triangle^{\mu\sigma}\nabla_\sigma u^{\nu}+\triangle^{\nu\sigma}\nabla_\sigma u^{\mu}-\tfrac{2}{3}\triangle^{\mu\nu}\nabla_\sigma u^{\sigma})$. Here $\triangle^{\mu\nu}=g^{\mu\nu}+u^\mu u^\nu$ is the standard three-frame projector and $\eta=\eta_0 \tau^{-1}$ is the shear viscosity of the plasma. Further coefficients in (\ref{e.aexp}) were computed in \cite{HigherOrder}, and are found to encode higher order dissipative hydrodynamics. We will focus on the metric at first order only, and leave the study of these higher order corrections to a more extensive report \cite{us}.

Before proceeding further, let us make some general remarks. For $v\to0$ the metric (\ref{metric}) reduces to pure AdS and, therefore, in this limit we expect to recover the known results for the various observables in the vacuum of the CFT \cite{Hubeny:2012ry}. According to the UV/IR connection \cite{uvir}, the bulk coordinate $z$ maps into a length scale $L\sim z$ in the boundary theory. Therefore, in the late-time regime $L\tau^{-\f{1}{3}}\epsilon^{\f{1}{4}}\to0$, we expect all the probes to relax to their corresponding AdS solution. Our goal is then to extract the leading order correction in the small parameter $\chi\equiv L\tau^{-\f{1}{3}}\epsilon^{\f{1}{4}}$, which is valid in the final stage of the evolution. Within this regime, we can explore the behavior as the other dimensionless parameter, $\xi\equiv\tau^{-\f{2}{3}}\eta_0$, is varied bearing in mind that $\xi$ must be small enough so that the hydrodynamical description still applies. Finally, note that for an expanding plasma there is no notion of thermodynamics given that we are dealing with an out-of-equilibrium configuration. However, at late times, the system can be consider near equilibrium and the energy density provides a definition of an \textit{effective} temperature
\be
T\equiv\left(8\varepsilon/(3\pi^2N_c^2)\right)^{\frac{1}{4}}\sim\tau^{-\frac{1}{3}}\epsilon^{\f{1}{4}}.
\ee
Thus, we have the gravity dual of a plasma undergoing cooling during expansion. 
The regime we are interested in corresponds to the low temperature regime $LT\ll1$. Hence, we are looking at approximate solutions in the same spirit as in \cite{Fischler:2012ca} for the case of a static plasma.

\noindent \textbf{3. Two-point correlators.}
The idea here is to study the thermal two-point function of an operator with large conformal
dimension. In this limit we can make use of the saddle point approximation and the problem reduces to the computation of geodesics \cite{Balasubramanian:1999zv}
\be
\langle \cO(t, \vec{x}) \cO(t', \vec{x}') \rangle \sim e^{- \Delta \cS_{\rm ren}}\,,
\ee
where $\Delta$ is the conformal dimension and $\cS_{\rm ren}$ is the renormalized geodesic length.

As a first step let us consider a spacelike geodesic connecting two boundary points separated in the transverse plane: $(\tau,x) = (\tau_0, -\tfrac{\Delta x}{2})$ and $(\tau', x') = (\tau_0, \tfrac{\Delta x}{2})$, where $x\equiv x_1$ and all other spatial directions are identical. Such a geodesic can be parametrized by two functions $\tau(z)$ and $x(z)$, with boundary conditions:
\be\label{realbc}
\tau(0) = \tau_0 \ , \quad x(0) = \pm \frac{\Delta x}{2} \ .
\ee
The action for this geodesic is given by
\be\label{actiongeo1}
\cS=2\int_{0}^{z_*}\f{dz}{z} \sqrt{1+e^{c}x'^2-e^{a}\tau'^2}\,.
\ee
Of course, in the strict limit $v\to0$ we expect to recover the action for a geodesic in AdS. Expanding around $v=0$, we get $\cS=\cS^{(0)}+\cS^{(4)}+\cO(v^8)$ where\footnote{Because $v=v(z)$ is a dynamical variable, we introduce a (dimensionless) scaling parameter through $v\to \zeta v$ and perform a Taylor expansion around $\zeta=0$. Then, we simply restore $\zeta=1$.}
\bea
&&\cS^{(0)}=2\int_{0}^{z_*}\!\f{dz}{z} \sqrt{1+x'^2-\tau'^2}\,,\nonumber\\
&&\cS^{(4)}=\frac{1}{3}\int_{0}^{z_*}\! \f{dz}{z} \frac{(x'^2+3 \tau'^2-6\eta_0\tau^{-\f{2}{3}}\tau'^2) v^4}{\sqrt{1+x'^2-\tau'^2}}\,.\nonumber
\eea
At zeroth order, it can be checked that the AdS solution $\tau(z)=\tau_0$ and $x(z)$ given in (\ref{soln}) (for $n=1$) is indeed a solution to the equations of motion derived from $\cS^{(0)}$. In this limit the renormalized geodesic length evaluates to (see the Appendix for details):
\be
\cS^{(0)}_{\text{ren}}=2\log L\,,\quad L\equiv\Delta x\,.
\ee
Taking into account the next-to-leading order term in the action changes the equations of motion for $\tau(z)$ and $x(z)$. However, it is easy to see that the corrections in these functions will only contribute at higher order in $v$ when the action is evaluated \textit{on shell}. Therefore, at our order of approximation it is still valid to evaluate $\cS^{(4)}$ in the AdS solutions. A brief calculation leads to
\be\label{s41}
\cS^{(4)}=\frac{1}{90}\,\chi^4\,,\quad \chi \equiv L \tau_0^{-\f{1}{3}}\epsilon^{\f{1}{4}}\,,
\ee
independent of the viscosity. Putting it all together, we find that the late-time behavior of the two-point correlator decays exponentially as
\be
\langle \cO(x_1)\cO(x'_1) \rangle \sim \frac{1}{|x_1-x_1'|^{2\Delta}}e^{-\frac{\epsilon\Delta|x_1-x_1'|^4}{90 \tau^{4/3}}}\,.
\ee

For longitudinal separation we can proceed in a similar way. In this case we are interested in a spacelike geodesic connecting the two boundary points $(\tau_0,x_3)$ and $(\tau_0, x_3')$. We can make use of the invariance under translations in $y$ and parameterize the geodesic by functions $\tau(z)$ and $y(z)$, with boundary conditions
\be\label{realbc}
\tau(0) = \tau_0 \ , \quad y(0) = \pm \frac{\Delta y}{2} \ .
\ee
At the end, we can simply shift our rapidity coordinate $y\to y+y_0$ and express our results in terms of
\be\label{y0defs}
x_3=\tau_0 \sinh (y_0+\tfrac{\Delta y}{2})\,,\quad x_3'=\tau_0 \sinh (y_0-\tfrac{\Delta y}{2})\,.
\ee
The action for this geodesic reads
\be\label{actiongeo2}
\cS=2\int_{0}^{z_*}\f{dz}{z} \sqrt{1+e^{b}\tau^2 y'^2-e^{a}\tau'^2}\,.
\ee
Expanding in $v$ we get $\cS=\cS^{(0)}+\cS^{(4)}+\cO(v^8)$, where
\bea
&&\cS^{(0)}=2\int_{0}^{z_*}\!\f{dz}{z} \sqrt{1+\tau^2 y'^2-\tau'^2}\,,\nonumber\\
&&\cS^{(4)}=\frac{1}{3}\int_{0}^{z_*}\! \f{dz}{z} \frac{[\tau^2 y'^2+3 \tau'^2-6\eta_0(\tau^{\f{4}{3}}y'^2+\tau^{-\f{2}{3}}\tau'^2)] v^4}{\sqrt{1+\tau^2 y'^2-\tau'^2}}.\nonumber
\eea
The term $\cS^{(0)}$ is the action of a geodesic in AdS written in proper time and rapidity coordinates. It is straightforward to check that the following embedding is a solution at zeroth order:
\be
\tau(z)=\sqrt{t_0^2-x(z)^2}\,,\quad y(z)=\mathrm{arccosh}\left(\frac{t_0}{\tau(z)}\right)\,,
\ee
where $t_0$ is a constant and $x(z)$ is the function given in (\ref{soln}) (for $n=1$). The relation between $(t_0,\Delta x_3)$ and $(\tau_0,\Delta y)$ is\footnote{More in general $\Delta x_3=2\tau_0\cosh (y_0) \sinh (\tfrac{\Delta y}{2})$ for $y_0\neq0$.}
\be
t_0=\tau_0 \cosh (\tfrac{\Delta y}{2})\,,\quad\Delta x_3= 2\tau_0 \sinh (\tfrac{\Delta y}{2})\,.
\ee
Of course, at zeroth order we have translation invariance in $x_3$ and the on-shell action reduces to
\be
\cS^{(0)}_{\text{ren}}=2\log L\,,\quad L\equiv\Delta x_3\,.
\ee
In this case, the first correction to the action is
\be
\cS^{(4)}=f(\Delta y,\xi)\,\chi^4\,,\quad \chi \equiv L \tau_0^{-\f{1}{3}}\epsilon^{\f{1}{4}}\,,
\ee
where $f(\Delta y,\xi)$ is given by the dimensionless integrals
\bea
&&f(\Delta y,\xi)=\int_0^1 \frac{x^5[3 x^2+(4-3 x^2) \cosh(\Delta y)-2]}{96(1-x^2)^{1/2}[x^2 \sinh^2(\f{\Delta y}{2}) + 1]^{5/3}}dx\nonumber\\
&&\qquad\qquad\,\,-\xi\int_0^1\frac{x^5[x^2+(2-x^2)\cosh(\Delta y)]}{16(1-x^2)^{1/2}[x^2 \sinh^2(\f{\Delta y}{2}) + 1]^{2}}dx\nonumber
\eea
and $\xi\equiv\tau_0^{-\f{2}{3}}\eta_0$. Finally, the two-point correlator for longitudinal separation evaluates to
\be
\langle \cO(x_3)\cO(x'_3) \rangle \sim \frac{1}{|x_3-x_3'|^{2\Delta}}e^{-\frac{\epsilon\Delta|x_3-x_3'|^4f(\Delta y,\xi)}{\cosh^4(y_0)\tau^{4/3}}}\,,
\ee
which is manifestly not invariant under translations.
\begin{figure}[t]
\begin{center}
\includegraphics[scale=0.9]{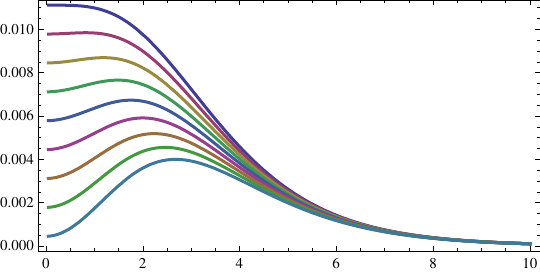}
\end{center}
\vspace{-0.7cm}
\caption{Contours of $f(\Delta y,\xi)$ for fixed $\xi=\{0,...,0.16\}$ with increments of $0.02$, from top to bottom, respectively.\label{fig1}}
\vspace{-0.3cm}
\end{figure}
The function $f(\Delta y,\xi)$ can be evaluated in terms of hypergeometric functions but we refrain from writing out the explicit result here since it is not particularly illuminating. In Fig. \ref{fig1} we plot $f$ as a function of $\Delta y$ for some fixed values of $\xi$. A few comments are in order. First, notice that for a fixed value of $\Delta y$, $f$ decreases as $\xi$ is increased. This is consistent with the fact that the viscosity damps the dynamics of the plasma and, therefore, we expect slower decorrelation as $\xi$ increases. For $\xi=0$, the function is monotonically decreasing in $\Delta y$, and interpolates from the $1/90$ coefficient found for the transverse case (\ref{s41}) to $f(\Delta y,\xi)\sim\cO(e^{-\frac{2\Delta y}{3}})\to0$ at large $\Delta y$. For finite $\xi$, the function is nonmonotonous and the small $\Delta y$ behavior is modified to
\be\label{smalldy}
f(\Delta y,\xi)= \frac{1-6\xi}{90}+\cO((\Delta y)^2)\,.
\ee
Note that $\xi$ has to be small in order for the hydrodynamic description to be valid. Indeed, from (\ref{smalldy}) we can already see that the approximation breaks down when $\xi>1/6$, in which case the function $f(\Delta y,\xi)$ flips sign and the vacuum value is reached from below.

\noindent \textbf{4. Wilson loops and entanglement entropy.}
Wilson loops and entanglement entropy are another two relevant sets of nonlocal observables. The Wilson loop operator is a path ordered contour
integral of the gauge field, defined as
\be
W(\mathcal{C})=\frac{1}{N_c}\mathrm{tr}\left(\mathcal{P}e^{\oint_\mathcal{C} A}\right)\,,
\ee
where the trace runs over the fundamental representation and $\mathcal{C}$ is a closed loop in spacetime. In AdS/CFT, the recipe for computing the expectation value of a Wilson loop, in the strong-coupling limit, is given by \cite{Wloops}
\be
\langle W(\mathcal{C}) \rangle = e^{-\mathcal{S}_{\text{NG}}(\Sigma)}\,,
\ee
where $\mathcal{S}_{\text{NG}}=(2\pi\alpha')^{-1}\times\mathrm{Area}(\Sigma)$ is the Nambu-Goto action and $\Sigma$ is an extremal surface with $\p \Sigma=\mathcal{C}$.

Entanglement entropy of a region $A$ with its complement $B$ is defined as the von Neumann entropy,
\begin{equation}
S_A = - {\rm tr}_A \, \rho_A \log \rho_A \,,
\end{equation}
where $\rho_A$ is the reduced density matrix, $\rho_A = {\rm tr}_{B}(\rho)$.
In AdS/CFT, the recipe for the computation of entanglement entropy is given by \cite{EEinAdSCFT},
\be \label{hrt}
S_A = \frac{1}{4 G_{\text{N}}} {\rm Area} \left(\gamma_A \right) \ ,
\ee
where $G_N$ is the bulk Newton's constant, and $\gamma_A$ is an extremal surface such that $\partial \gamma_A = \partial A$.

From the point of view of the bulk, these two sets of nonlocal observables are natural generalizations of the two-point functions considered above. In particular, they involve the computation of extremal surfaces of higher dimensions and therefore, the results of the Appendix (for $n=2,3$) will now become handy. We will only deal with some loop contours $\mathcal{C}$ and regions $A$ that can be treated analytically in the same way as the two-point correlators. For such cases, the computations proceed identically as before so we will just state the final results. A more detailed explanation (and the results for other shapes) will be given elsewhere \cite{us}.

For the Wilson loop we consider two cases. The first case consists of a rectangular loop in the transverse plane, where $x_1\in[-\frac{\Delta x}{2},\frac{\Delta x}{2}]$, $x_2\in[-\frac{\ell}{2},\frac{\ell}{2}]$ and $\ell\to\infty$. We will call this case $W_\perp$. The second case is also a rectangular loop but in this case one side is along the longitudinal direction, $y\in[-\frac{\Delta y}{2},\frac{\Delta y}{2}]$, $x_1\in[-\frac{\ell}{2},\frac{\ell}{2}]$ and $\ell\to\infty$. We will refer to this one as $W_\parallel$. For the first case we find an exponential decay
\begin{figure}[t]
\begin{center}
\includegraphics[scale=0.9]{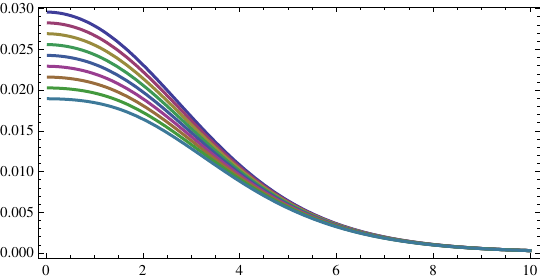}
\end{center}
\vspace{-0.7cm}
\caption{Contours of $g(\Delta y,\xi)$ for fixed $\xi=\{0,...,0.16\}$ with increments of $0.02$, from top to bottom, respectively.\label{fig2}}
\vspace{-0.3cm}
\end{figure}
\be
\langle W_\perp\rangle\sim\langle W^{(0)}\rangle e^{-\frac{\epsilon\sqrt{\lambda}\ell(\Delta x_1)^3\Gamma(1/4)^4}{60 \pi^4 \tau^{4/3}}}\,,
\ee
where $\langle W^{(0)}\rangle\equiv\exp[\frac{4\pi^2\sqrt{\lambda }\ell}{\Gamma(1/4)^4 \Delta x_1}]$ is the vacuum expectation value of the Wilson loop. For the second case, on the other hand, we find that the exponential is now modulated by a function $g(\Delta y,\xi)$,
\be
\langle W_\parallel\rangle\sim\langle W^{(0)}\rangle e^{-\frac{\epsilon\sqrt{\lambda}\ell(\Delta x_3)^3g(\Delta y,\xi)}{\cosh^3(y_0)\tau^{4/3}}}\,.
\ee
In the regime $\Delta y\ll1$, the function $g$ behaves as
\be\label{smalldy2}
g(\Delta y,\xi)= \frac{\Gamma(\frac{1}{4})^4}{240 \pi ^4}(4-9 \xi)+\cO((\Delta y)^2)\,,
\ee
which imposes the (weaker) constraint $\xi<4/9$. At large $\Delta y$, we find $g(\Delta y,\xi)\sim\cO(e^{-\frac{2\Delta y}{3}})$. Some contours of $g$ for fixed values of $\xi<1/6$ are given in Fig. \ref{fig2}. Notice that, contrary to the two-point function, in this range the function $g$ always decreases monotonically in $\Delta y$.

\begin{figure}[t]
\begin{center}
\includegraphics[scale=0.9]{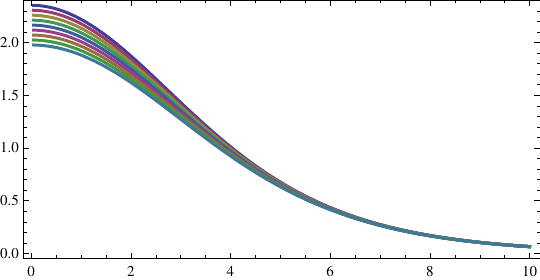}
\end{center}
\vspace{-0.7cm}
\caption{Contours of $h(\Delta y,\xi)$ for fixed $\xi=\{0,...,0.16\}$ with increments of $0.02$, from top to bottom, respectively.\label{fig3}}
\vspace{-0.3cm}
\end{figure}

For entanglement entropy we only consider the case where the region $A$ is a 3-dimensional strip with $y\in[-\frac{\Delta y}{2},\frac{\Delta y}{2}]$, $x_i\in[-\frac{\ell}{2},\frac{\ell}{2}]$ ($i=1,2$) and $\ell\to\infty$. A brief computation leads to
\be
S_A=S_A^{(0)}\left(1-\frac{\epsilon(\Delta x_3)^4h(\Delta y,\xi)}{\cosh^2(y_0)\tau^{4/3}}\right)\,,
\ee
where $S_A^{(0)}\equiv-\frac{2\sqrt{\pi}\Gamma(\f{4}{6})^3\ell^2N_c^2}{\Gamma(\f{1}{6})^3(\Delta x_3) ^2}$
is the entanglement entropy of region $A$ in the vacuum state. The function $h(\Delta y,\xi)$ behaves in a similar way as $g(\Delta y,\xi)$, with slightly less sensitivity with respect to the viscosity. For small $\Delta y$ we find
\be\label{smalldy3}
h(\Delta y,\xi)= \frac{\Gamma(\frac{1}{6})^9}{1280 \pi ^{13/2}}(1-\xi)+\cO((\Delta y)^2)\,,
\ee
whereas for large $\Delta y$, $h(\Delta y,\xi)\sim\cO(e^{-\frac{2\Delta y}{3}})$. The constraint on the viscosity is even weaker in this case: $\xi<1$, which is due to the fact that in any excited state we expect a higher entanglement than in the vacuum. Some contours of $h$ for fixed $\xi$ are given in Fig. \ref{fig3}.

\noindent \textbf{5. Discussion.}
We have studied the late-time behavior of various nonlocal observables in a maximally supersymmetric Yang-Mills plasma undergoing a boost-invariant expansion, including effects of first order dissipative hydrodynamics. More specifically, we gave analytic expressions for the evolution of certain two-point functions, Wilson loops and entanglement entropies in the regime $\chi=L\tau^{-\f{1}{3}}\epsilon^{\f{1}{4}}\ll1$ (where $L$ is the typical size of the probe) as a function of $\xi=\tau^{-\f{2}{3}}\eta_0$.

The two-point correlators and Wilson loops are found to relax exponentially in $\chi$ whereas the entanglement entropy equilibrates at a much slower rate (as a power of $\chi$). Hence, it is the entanglement that sets the relevant time scale for the approach to equilibrium, which is consistent with the known results of holographic thermalization \cite{thermalization}. Another interesting result is the dependence of the observables on the longitudinal variables. We find that, in such cases, the leading behavior of the observables is modulated by functions of $\Delta y$ and $\xi$. For fixed $\Delta y$, these functions decrease monotonically in $\xi$, which is consistent with the dissipative nature of the viscosity. For fixed $\xi$ we also find a monotonic behavior for the Wilson loop and entanglement entropy, but not for the two-point function. This is interpreted as a nontrivial prediction from AdS/CFT. Indeed, one would naively expect a higher correlation for points with similar rapidities, which does not hold for some range of the parameter space. Finally, it is worth pointing out that our results point to a breakdown of the first order hydrodynamics for $\xi>1/6$, which is set by the behavior of the longitudinal two-point function at small $\Delta y$.

\vskip .2cm
\noindent{\bf Acknowledgments.}
I am grateful to E.~C\'aceres, B.~S.~DiNunno, W.~Fischler, D.~Giataganas, A.~G\"uijosa, M.~Paulos and S.~Young for useful conversations and comments on the manuscript. I also want to thank the organizers and participants of Mexicuerdas/Mextrings 2014, especially to L.~Pando-Zayas, E.~Tonni and D.~Trancanelli for the feedback provided and the various discussions on related topics. This material is based upon work supported by the National Science Foundation under Grant No. PHY-1316033 and by the Texas Cosmology Center, which is supported
by the College of Natural Sciences and the Department of Astronomy at the University
of Texas at Austin and the McDonald Observatory.

\vskip .2cm
\centerline{\bf Appendix: Extremal surfaces in AdS$_5$}
\vskip .2cm
Following \cite{Hubeny:2012ry}, we review the extremal surfaces in AdS$_5$ ending on a strip. In the Poincare patch the metric is given by
\be
ds^2=\frac{1}{z^2}\left( -dt^2+d\vec{x}^2 +dz^2\right)\,.
\ee
We define a boundary region $A$ to be an $n$-dimensional strip with $x_1\in[-\frac{\Delta x}{2},\frac{\Delta x}{2}]$ and $x_i\in[-\frac{\ell}{2},\frac{\ell}{2}]$ for $i=2,...,n$. We assume $\ell\to\infty$, so there is translation invariance along the transverse directions. We want to find the extremal surface $\Gamma_A$ in the bulk that is anchored on $\p A$. Choosing the coordinates on $\Gamma_A$ to be $\sigma^1=x_1\equiv x$ and $\sigma^i=x_i$ for $i=2,...,n$ and parametrizing the surface by a single function $z(x)$ we get that the area functional is given by
\be\label{arean}
\mathcal{A}\equiv \mathrm{Area}(\Gamma_A)=\ell^{n-1}\int_{-\frac{\Delta x}{2}}^{\frac{\Delta x}{2}}\frac{\sqrt{1+z'^2}}{z^n}dx\,.
\ee
Since there is no explicit dependence on $x$, the Hamiltonian is conserved,
\be
\mathcal{H}=\frac{\p \mathcal{L}}{\p z'}z'-\mathcal{L}=\frac{-1}{z^n\sqrt{1+z'^2}}\equiv\frac{-1}{z_*^n}\,,
\ee
where $z_*$ is defined through $x(z_*)=0$. This allows us to obtain an explicit expression for $x(z)$:
\be\label{soln}
\pm x(z)=\frac{\Delta x}{2}-\frac{z^{n+1}}{(n+1) z_*^n}\!\,_2F_1
\left[\tfrac{1}{2}, \tfrac{n+1}{2n},  \tfrac{3n+1}{2n}, \tfrac{z^{2n}}{z_*^{2n}} \right],
\ee
from which we can obtain
\be
z_*= \frac{n \Gamma( \frac{2n+1}{2n}) }{\sqrt{\pi} \Gamma( \frac{n+1}{2n}) } \Delta x \ .
\ee
Finally, the area of the extremal surface can be computed evaluating (\ref{arean}) on shell,
\be\label{onshell}
\mathcal{A}=2\ell^{n-1}\int_{z_0}^{z_*}\frac{dz}{z^n\sqrt{1-(z/z_*)^{2n}}}\,.
\ee
This quantity is UV divergent given that in the limit $z_0\to0$ the surface reaches the boundary of AdS. The divergent piece can be isolated by studying the near-boundary behavior of (\ref{onshell}):
\begin{align}
 \mathcal{A}_{\mathrm{div}}=2\ell^{n-1}\int_{\!z_0}\! \frac{dz}{z^n}=
 \begin{cases}
  \displaystyle -2\log z_0\,, & \displaystyle n=1 \,,\\[1ex]
  \displaystyle \frac{2\ell^{n-1}}{(n-1)z_0^{n-1}}\,,  & \displaystyle n>1 \,.
 \end{cases}
\end{align}
Subtracting this divergence, we obtain the finite term which is the main
quantity we are interested in:
\begin{align}
\mathcal{A}_{\textrm{ren}}=
 \begin{cases}
 \displaystyle 2\log \Delta x\,, & \displaystyle n=1 \,,\\[0.8ex]
  \displaystyle -\frac{8\pi^3}{\Gamma(\f{1}{4})^4}\frac{\ell}{\Delta x}\,,  & \displaystyle n=2 \,,\\[2ex]
  \displaystyle -\frac{4\pi^{3/2}\Gamma(\f{4}{6})^3}{\Gamma(\f{1}{6})^3}\frac{\ell^2}{(\Delta x) ^2}\,,  & \displaystyle n=3 \,.
  \end{cases}
\end{align}

\end{document}